# Image Understanding and the Web: a State-of-the-Art Review

Fariza Fauzi
Department of Engineering Science
University of Auckland, New Zealand

Mohammed Belkhatir
Faculty of Computer Science
University of Lyon, France

**ABSTRACT**
The contextual information of Web images is investigated to address the issue of characterizing their content with semantic descriptors and therefore bridge the semantic gap, i.e. the gap between their automated low-level representation in terms of colors, textures, shapes… and their semantic interpretation. Such characterization allows for understanding the image content and is crucial in important Web-based tasks such as image indexing and retrieval. Although we are highly motivated by the availability of rich knowledge on the Web and the relative success achieved by commercial search engines in automatically characterizing the image content using contextual information in Web pages, we are aware that the unpredictable quality of the contextual information is a major limiting factor. Among the reasons explaining the difficulty to leverage on the image contextual information, some problems are related to the characterization and extraction of this information. Indeed, the first issue is the lack of large-scale studies to highlight what is considered the relevant contextual information of an image, where it is located in a Web page and whether it is consistent across Web pages of different types, content layouts and domains. Also, the matter related to the extraction of this contextual information is topical as state-of-the-art automated extraction tools are unable to handle the heterogeneous Web. As far as the processing of the contextual information is concerned, problems linked to the syntactic and semantic characterizations of the textual components are important to address in order to tackle the semantic gap. Furthermore, questions pertaining to the organization of these textual components into coherent structures that are usable in image indexing and retrieval frameworks shall arise. To address these issues, we lay down the anatomy of a generic context-based Web image understanding framework and propose its stage-based decomposition, covering topical issues from information indexing and retrieval, image description models, natural language processing, Web page segmentation and automated information extraction. For each of the identified stages, we review state-of-the-art solutions in the literature categorized and analyzed under the light of the techniques used.

## 1. INTRODUCTION

In the recent years, there has been a dramatic increase in the number of images due to advance storage technology as well as affordable digital cameras. Furthermore, we have the World Wide Web (shortened as the Web) making these images accessible on a global scale. The Web has transformed the way people communicate and socialize. The new wave of social networking websites such as Facebook, MySpace, Flickr, YouTube etc. are enabling people to upload and share images and other multimedia contents. Thus, the Web can be viewed as an inexhaustible collection of images across diverse domains and usually, these images come with rich contextual information or surrounding information (the terms are used interchangeably throughout this paper), which is the text associated to the images, used jointly with their filename, alt description, and page title.

Current commercial Web-based image search engines such as Google Image, Yahoo!, Bing, etc. automatically index images using this information in their keyword-based image retrieval systems. These systems work very well for most simple or general queries (e.g. cat, dog, Michael Jackson, Twilight, etc.). However, for longer and more complex queries, in particular those featuring relational information such as "baby boy in blue" and "car hits lorry", the retrieved results are not as satisfactory since the information need conveyed is more specific. Such queries are common and the reason behind most of the failed searches (Pu, 2008; Chung and Yoon, 2010). This is shown in Figure 1.

The retrieval precision of these systems is highly dependent on the quality of the image contextual information. Not all entities in the image contextual information are related to the image and in particular, as far as the textual content is concerned, some words are inter-related for them to be meaningful to the image. An in-depth understanding of this information is the focus of this paper.

## 1.1 Image Understanding in the Context of Image Indexing & Retrieval Systems

Image retrieval systems are defined as computer systems for searching and retrieving images from large collections of digital images according to the users' information needs. Examples of image corpora include medical images, satellite images, photo galleries, the Web, etc.

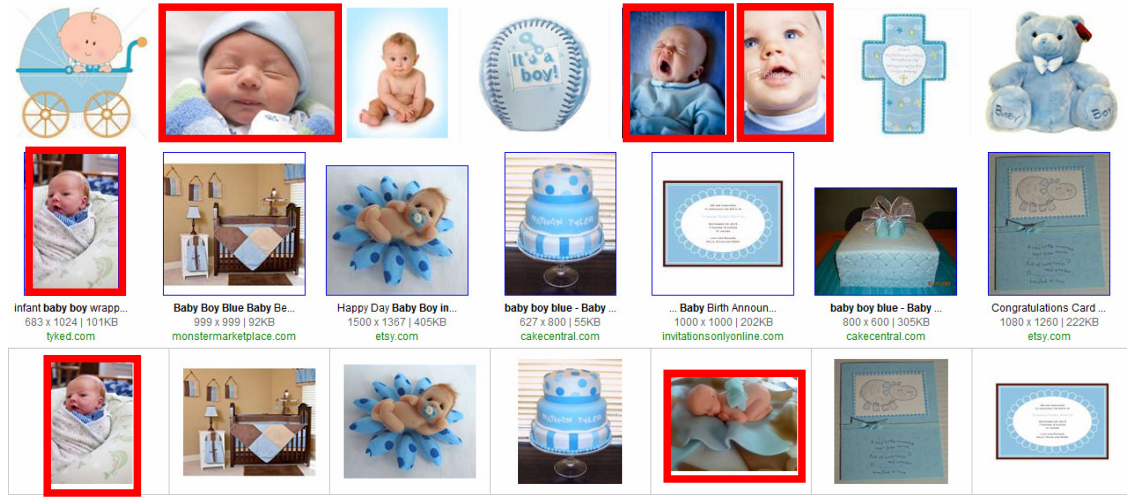

(a) The three rows shown are the first rows of the retrieved search results of the Google, Yahoo and Bing search engines respectively for the query "Baby boy in blue". 6 images (framed in red) out of 22 images are relevant.

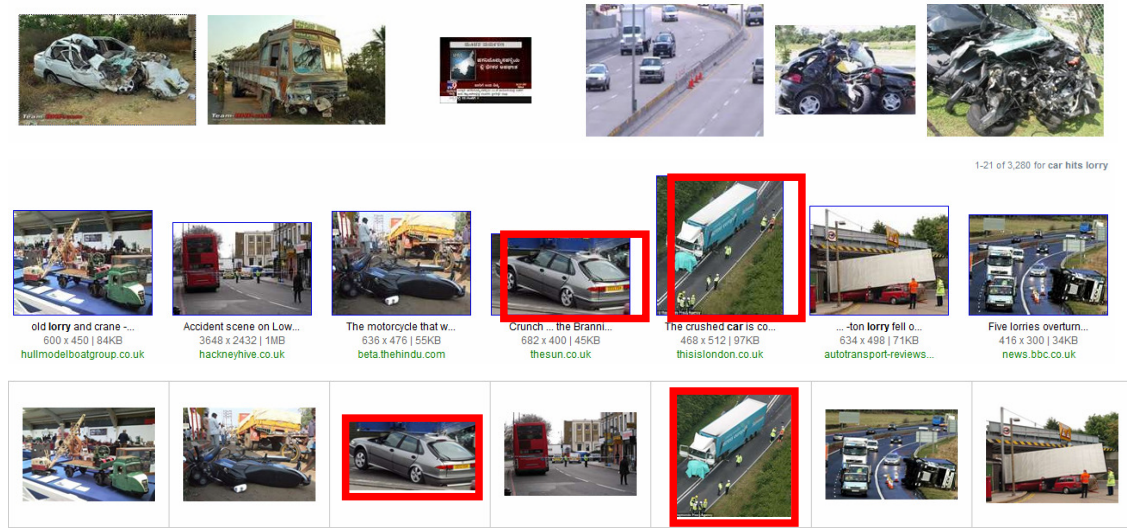

(b) The three rows shown are the first rows of the retrieved search results of the Google, Yahoo and Bing search engines respectively for the query "Car hits lorry". 4 images (framed in red) out of 20 images are relevant.

**Figure 1.** The retrieved results for two search queries from three current Web image search engines.

There are two aspects in an image retrieval system: indexing and searching. *Indexing* involves processing the image and representing it in some form and *searching* involves comparing the query, which conveys the user's need to the system, to the indexes of all the images in the collection. The images with indexes that match the query are returned to the user. Clearly, the performance of image retrieval systems is impacted by the ability to decipher the image content. A typical image retrieval system is depicted in Figure 2.

The traditional keyword-based systems involve the manual tagging of images with keywords, initially costly and impractical as the corpus size increases. However, with the Web 2.0 technology, manual tagging is achievable as implemented in Flickr, the online photo sharing website. Nevertheless, the drawback of such methods is still present that it is highly dependent on the indexer; the annotations are error-prone, not comprehensive and the range of successful queries is limited to the interpretation of the indexer/annotator (Hollink et al., 2004; Inoue, 2004; Smeulders et al., 2000).

Automated image understanding techniques to address the image semantic gap are important to better image indexing techniques and therefore enhance the expressiveness of query languages. In the example of Figure 2, the system can effectively process the information need (i.e. retrieving a ladybird on the petals) that is submitted as a query only if it is able to characterize the image content in terms of its semantics : both the physical entities (i.e. "ladybird" and "petals") and the relational fact that the ladybird is "on" the petals.

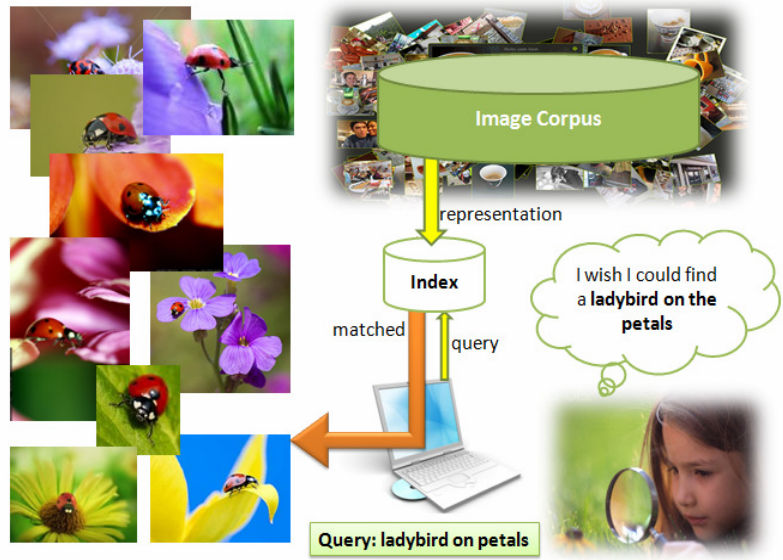

**Figure 2** Outline of an image retrieval system

## 1.2 Motivations for the need to investigate the Web context

Back to our focus on the WWW and its assorted image collections, there are many research works that incorporate the contextual information of Web images within image retrieval systems (Cai et al., 2004; Feng et al., 2004; Gao et al., 2005; Liu et al., 2009; Lu et al., 2000; Sclaroff et al., 1999; Wang et al., 2008). They draw on the contextual information as a source of countless semantic concepts at the same time attempting to address the poor retrieval performance of current Web-based systems with the image visual content.

Cai et al. (2004), Gao et al. (2005) and Wang et al. (2008) combine textual and visual information to cluster images rather than to annotate images. Our focus is on the fusion of the two sources for image annotation purposes as in the work of Sclaroff et al. (1999), Lu et al. (2000), Feng et al. (2004), Liu et al. (2009) and Wang et al. (2008). In the earlier papers (Lu et al., 2000; Sclaroff et al., 1999), the textual and visual information are kept in separate data stores and loosely coupled by means of relevance feedback where the user is required to provide feedback on which images are relevant (or irrelevant) out of all the images in his/her search result of their keyword-based query, making such systems less attractive to users. Feng et al. (2004) and Liu et al. (2009) strongly couple both textual and visual image contents using a bootstrapping approach and graph learning method, respectively. They improve the system recall rate (i.e. a measure of the ability of a system to present all relevant images) and effectively remove the need for user feedback. However, only a minor increase is obtained in the system retrieval precision (i.e. a measure of the ability of a system to present only relevant images). A possible reason for this is the inaccuracy of the textual information attached to the image, which brings us back to the original problem of unpredictable quality of the contextual information of a Web image faced by the current Web-based systems.

This makes the need to understand the contextual information of a Web image crucial. A review of the literature reveals two main problems affecting the quality of the image contextual information. Firstly, there is no straightforward process to define the locations of relevant contextual information. While the image filename, alt description and page title are generally accepted by researchers in the area, the difference is in the contextual textual content, which has been perceived either as: i) a window of words (http://image.google.com; Sclaroff et al., 1999; Coelho et al. 2004); ii) a paragraph (Frankel et al., 1996; Mukherjea et al., 1999; Shen et al., 2000); iii) a section/segment (Cai et al., 2004; Feng et al., 2004; Hua et al., 2005; Li et al., 2006) and; iv) even the entire page (Gao et al., 2005; Gong et al., 2006; Ortega-Binderberger et al, 2000; Rege et al., 2008). The differences of opinions amongst researchers can be attributed to the fact that Web page partitioning is a difficult problem due to the diverse page layouts. Secondly, there is a lack of semantic techniques in the indexing process. Standard text processing techniques are commonly employed to filter out irrelevant words (Feng et al., 2004; Gao et al., 2005; Hua et al., 2005; He et al., 2007; Liu et al., 2009). These techniques, which include stemming, stop-word removal, retention of noun phrases, frequent words and/or co-occurring words, etc. are typically employed in a "bag-of-words" representation model. While these techniques adequately address short and generic queries, they perform poorly on longer or specific queries. Nevertheless, we cannot ignore the longer or specific queries as user studies on user queries have shown that such queries are common (Armitage and Enser, 1997; Hollink et al., 2004; Pu, 2008; Chung and Yoon, 2010). An image retrieval system should therefore be able to cater to both short/generic and long/specific queries. Thus, more knowledge of the characteristics of the relevant image contextual information is required for optimal usage.

## 1.3 Research Aims and Objectives

The heterogeneity of Web page designs and layouts as well as the non-conformity of authoring Web page content are non-trivial problems. In this paper, we attempt to address these. We do not restrict ourselves to any specific domain or any specific type of Web page layout and content in the investigation of the contextual information of Web images; it is, however, limited to English content only. There are two sets of research questions that this paper aims to address. The first set consists of the following questions:

*Given an image on a Web page,*
- *What is considered the relevant contextual information of an image and where is it located?*
- *Is it the same for all kinds of Web pages?*
- *How can the contextual information be reliably and automatically extracted for any kind of Web page?*

The set of questions above is motivated by the varying definitions of image contextual information. Knowing the reasons behind the absence of a standard definition will assist us in our search for a solution that will adequately extract the information from any kind of Web page. Having obtained this information, finding the answers to the following set of questions will help us understand it further and identify mechanisms to discriminate between the relevant and irrelevant contextual information.

*Given an image with its surrounding information,*
- *Are all the link-based and textual components in its contextual information relevant to the image?*
- *What are the (syntactic and semantic) characteristics of the textual components?*
- *How do we reliably extract components that relevantly describe the image?*
- *How do we organize the set of extracted information into a coherent structure that is understood by users?*

A study on the available image description models, which highlight the importance of conceptual categories (Hearst, 2006; Hollink et al., 2004; Jaimes and Chang, 2000; Armitage and Enser, 1997; Yee et al., 2003), might provide insights on ways to structure the extracted information at the conceptual level that matches the users' information requirements. Some form of weighting scheme is needed as well to rank each word in the extracted information according to its relevance to the image content. Too much or too little indexes will affect the systems' performance in terms of precision and recall. Taking all terms in the image contextual information might improve the system recall performance as we would surely have included all the relevant terms but at the same time, it might also decrease the system precision if there are many irrelevant terms present.

Our goal eventually is to propose the anatomy of a context-based indexing framework that automates this task and utilizes the relevant image contextual information as indexes or annotations in a structured manner. We expect to improve the quality of the surrounding image information, which in turn will improve the system performance.

The rest of the paper is organized as follows: Section 2 identifies the important topics on the use of contextual information for image understanding on the Web. In Section 3, we propose a categorization of image retrieval systems using context-based image understanding techniques. In Section 4, we outline the anatomy of a context-based indexing framework that processes the relevant image contextual information into a coherent and structured semantic-based representation, also covering the requirements for the automation of this task. We present its decomposition into two stages of definition and extraction of relevant image contextual information on the one hand and semantic processing of this information on the other hand. In Sections 5 and 6, we discuss the usage of contextual information in the state-of-the-art in accordance to the two highlighted stages. We take a close look at issues within each stage and how they are being tackled in related works. In Section 5, the stage of definition and extraction encompasses establishing the amount and type of contextual information considered relevant with respect to the image content and performing the automatic extraction of this contextual information. In Section 6, we discuss the techniques used to convert the contextual information into an image representation based on semantic-based index terms (which could be words or phrases). Lastly, we deal with weighting techniques, which involve emphasizing the contribution of some features of index terms to estimate their relevance to the image content.

## 2. ISSUES IN CONTEXTUAL INFORMATION-BASED IMAGE UNDERSTANDING ON THE WEB

The topic of our paper centers on the textual information that surrounds or is attached to Web images. This contextual information is a unique feature of the Web images and has long been mined for various uses such as image annotation (Joshi and Liu, 2009; Leong at al., 2010; Li et al., 2006), clustering of image search results (Blaschko and Lampert, 2008; Cai et al., 2004; Gao et al., 2005; Rege et al., 2008; Wang et al., 2005), inference of image semantic content (Feng and Lapata, 2008; Ghoshal et al., 2005; Tang et al., 2009) etc.

The use of image contextual information is important due to user preference for text input query (Choi and Rasmussen, 2003; Hughes et al., 2003; Rorissa, 2008), which signifies the need for the images to be annotated/indexed with text. Furthermore, the latest trend of multi-modal systems (Liu et al., 2009; Rege et al., 2008; Wang et al., 2008), which fuse both visual and textual features of Web images, has demonstrated the need for better text processing techniques other than the typical stop word removal and stemming as the inaccuracy of the textual information attached to the Web images degrades their performance.

One main drawback of the contextual information is that it is subjective to the Web authors' (i.e. annotators') point of view, knowledge, culture, and experience (Kherfi et al., 2004; Inoue, 2009; Rorissa, 2010). Nevertheless, the contextual information is still rich with high-level semantic concepts (which are still hard to derive from content-based systems). Contextual information contains both direct and indirect information of the image e.g. the immediate objects, attributes, high-level knowledge, events etc. Having this contextual information available is definitely an advantage but, of course, its usage is not without problems and challenges.

Various problems can be identified starting from *defining* the locations of relevant image contextual information, which in turn influences the extraction techniques, the current *text processing* techniques employed to convert it into a set of index terms to its application in Web Image Retrieval Systems as far as *representation* (i.e. how is it used?) and *noise filtering* (i.e. how is the relevant information distinguished from the irrelevant information?) are concerned. The handling of these four stages is of importance when developing image retrieval systems that incorporate image contextual information.

The major difficulty in defining (and extracting) this information is due to the diverse Web page layouts and designs found in today's Web pages. Unlike the earlier editions of Web pages, Web pages currently contain various contents such as navigation, decoration, interaction, contact information and of course the main content. The main content of the modern Web page more often contains multiple topics that are not necessarily related to one another. Thus, the boundaries of the contextual information for each image on a Web page vary and considering a fixed number of words before and after an image as commonly implemented by some image retrieval systems (e.g. http://image.google.com; Coelho et al., 2004; Sclaroff et al., 1999) is questionable. Figure 3 illustrates the weakness of taking a static amount of text as the surrounding information of an image for such pages. It is a challenge indeed to acquire the right amount of information automatically without any human intervention and given any Web page layout.

The image contextual information generally consists of free text written by Web authors. For most parts, there is no governing body to control the content or monitor the quality of this information. It will be written and structured according to the Web authors' preferences. As such, converting this information into a suitable set of annotations is another difficult task. Typically, the extracted text is processed by breaking the text down into individual words, called tokens in a tokenization process and removing stop words. The remaining tokens are used to annotate the image in a bag-of-words representation model where the contextual information is represented as an unordered collection of terms to be used directly as index terms or as the textual feature vector in a multi-modal image retrieval system. Various weighting functions which mainly incorporate the term frequency feature have been used to estimate a term's importance to the image content. Lower ranked terms are filtered out as noise.

Such approach brings about the issue related to the selection of terms – is it appropriate to simply split the contextual information into individual words? What will happen to terms such as "fountain pen", "Tiger Balm", "heart attack"…? What about removing stop words which are generally defined as functional words such as articles, prepositions or conjunctions? This would pose problems for the following example terms: "bird <u>on</u> house" and "Cat <u>on</u> mat". The terms "bird on house" and "bird house" (as well as "cat on mat" and "cat mat") obviously differ semantically but for systems that simply perform stop word removal, they are considered the same as illustrated in Figure 4.

The bag-of-words representation limits the user queries to short and simple queries. Such representation is the main reason for the drawback of existing Web image search systems illustrated in Figure 1 whereby images of a blue monkey and a bear in blue are returned for the first query "*baby boy in blue*". We further highlight this weakness in Figure 5. The zoomed-in image shown in Figure 5 is returned in the first page of the search results for the query *"car hits lorry"* query. From a user's view, this image is irrelevant but to a typical keyword-based image retrieval system, as long as the word *car* exists in the image contextual information, the image is deemed as relevant and returned in the search result list. The associations between the "cow" and "car" from the image contextual information as well as "car" and "lorry" from the query are lost in a bag-of-words representation.

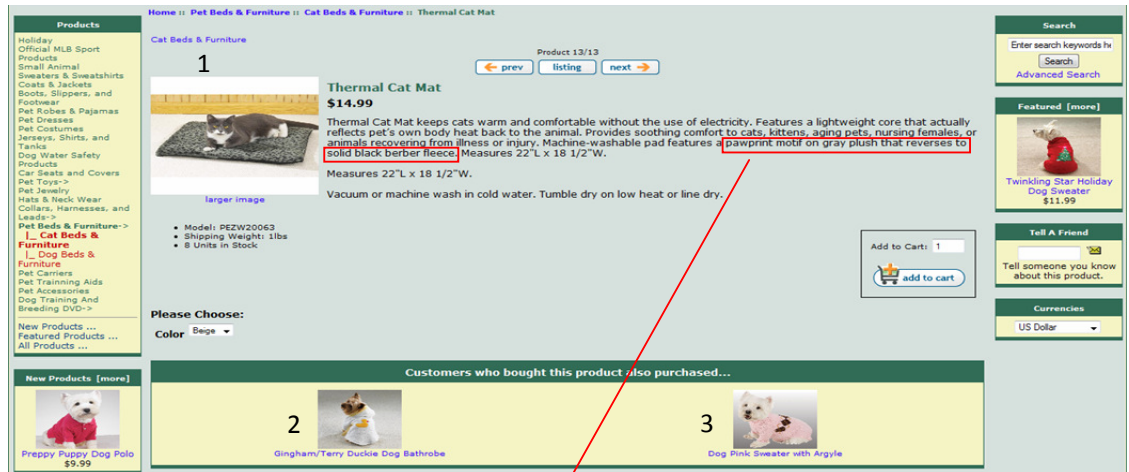

(a) A random Web page with URL: www.spoilurpets.com/thermal-cat-mat-p-150.html

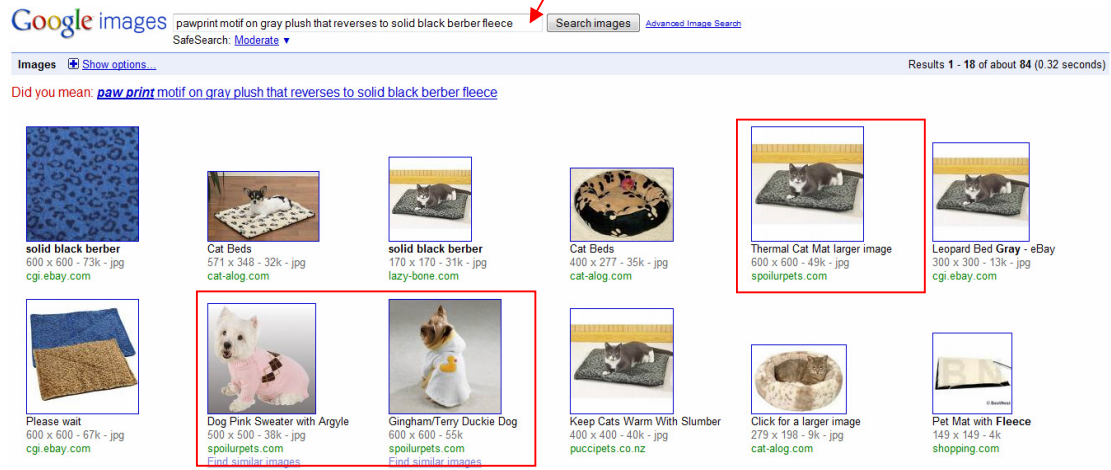

(b) The first page of the search results for the query "pawprint motif on gray plush that reverses to solid black berber fleece" shown by the Google search engine

**Figure 3.** An example: the usage of surrounding text by Google Image to annotate image. The surrounding terms highlighted in the red box in (a) are used as the query in (b) and we can see images wrongly indexed with those terms.

Another critical issue of using contextual information lies in the presence of noise (Kherfi et al. 2004; Inoue, 2009). This is addressed by estimating the relative importance of a given term as an image descriptor. Typically, in Information Retrieval, one of the common and effective methods is to give different terms different weights (Jones, 1973). Each term is assigned a *local weight*, indicating its importance within the particular document, as well as a *global weight*, indicating its overall importance in the collection as an index term. Therefore, the value for a term in a document is the product of the local and global weights. Popular local weightings include: *Term Frequency*, *Binary*, *Log* (*term frequency* + 1), etc. Four well-known global weightings are: *Normal*, *Gfidf*, *Idf* and *Entropy* (Dumais, 1991). The calculations of these weights typically depend on the frequency of a term. This poses problems for image with little contextual information. Other features of a term should be considered as well.

Nevertheless, even with all these issues, the contextual information is still considered as an important feature as we can see from the various contextual information-based image retrieval systems found in the literature. In the next section, a selection of these systems which explicitly describe their use of contextual information and are unique in their approach of using the contextual information is discussed. Our intention is to draw attention to the different implementations of image contextual information in existing state-of-the-art systems and not to present a broad survey on Web Image Retrieval Systems. Further readings on these systems in general can be found in (Datta et al., 2008; Kherfi et al., 2004; Liu et al., 2007).

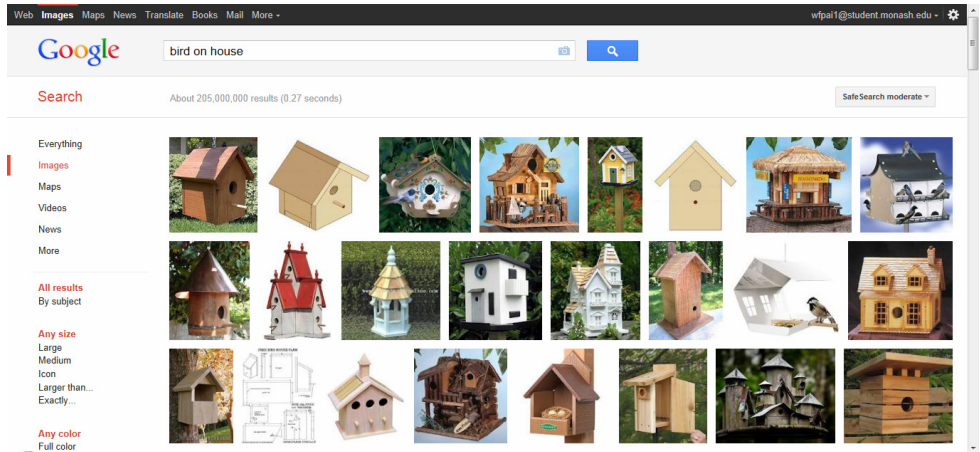

**Figure 4.** Search result for the query "Bird on house"

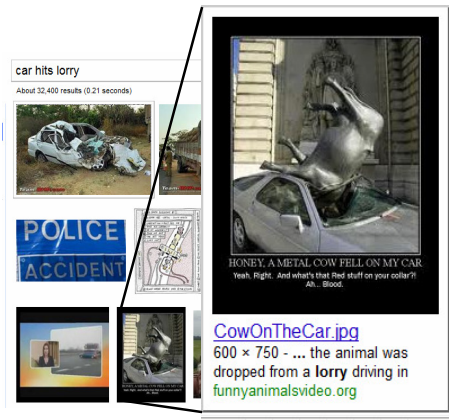

**Figure 5.** Search result for query "Car hits lorry"

## 3. IMAGE RETRIEVAL SYSTEMS USING CONTEXT-BASED IMAGE UNDERSTANDING TECHNIQUES ON THE WEB: A CATEGORIZATION

Image retrieval systems from the Web using context-based image understanding techniques can be categorized into two types: pure text-based systems or fused models (multi-modal systems). For each type of system, we analyze the usage of the image contextual information in the state-of-the-art.

### 3.1 Text-based Systems

When we talk about text-based Web image retrieval systems, the commercial image search engines – Google Image, Yahoo! Image and Bing Image (formerly Live Search Image and MSN Image) come to mind. They are the three major image search providers of our time and have originated from general text information search engines. They typically rely on text to index Web images. These keyword-based systems utilize the image filename, hyperlink text pointing to the image, and/or text adjacent to or surrounding the image. It is only recently in 2009 that Google included an image content-based feature – similarity feature, to lookup other similarly colored or textured images after the initial textual query (Evans, 2009). This feature is similar to the early fused concept-based models that loosely-couple textual and visual features where users can initially submit a text query and go on to select and view other visually similar images from the query result (Lew, 2000; Lu et al., 2000; Sclaroff et al., 1999; Smith and Chang, 1997). In 2009 as well, Yahoo! Search got powered by Bing Search including Yahoo! Image Search in a Microsoft and Yahoo Web deal (BBC News, 2009).

While it is hard to determine how much text these search engines consider as adjacent to an image, Feng et al. (2004) report that the first or last 32 words in the text nearest to an image appear to be the most descriptive of the image content according to a survey conducted by Google. For Bing Image Search, which is the latest launch from Microsoft, we assume that they have moved from considering a number of terms as the image surrounding text to a section of text extracted using their patented Web page segmentation algorithm, called VIPS, which partitions a Web page into several smaller semantic blocks (Slawski, 2008).

In Coelho et al. (2004), Gong et al. (2006), Joshi and Liu (2009), Leong and Mihalcea (2009), Leong et al. (2010) and Shen et al. (2000), examples of pure text-based systems are described. Shen et al. (2000) introduce an image representation model called *Weight ChainNet*. Weight ChainNet is based on a *Lexical Chain* (LC) that represents the semantics of an image from its nearby text. They consider the image filename, image ALT, page title and image caption (i.e. the entire paragraph containing the image) as the image contextual information and model these texts as different LCs in the Weight ChainNet Model. An LC is built as a sequence of semantically related words in a text. Each LC is heuristically weighted based on the text's representativeness (semantic relationship) to the image. These weighted LCs contribute to the semantic similarity computation in their proposed retrieval model. A proper combination of the LCs, each with its own appropriate weight, is shown to perform best. The optimized weight for each LC is obtained through several tests of different weight combinations. Query results are further refined via relevance feedback methods.

Coelho el al. (2004) consider image contextual information as text from multiple sources as well, with each part of the text being regarded as an independent source of evidential information. Four possible sources of evidence are proposed: description tags, meta tags, full text and text passages (i.e. surrounding text – words located close to the images). *Description tags* are words found in the image filename, the ALT attribute of the  tag and between the anchor tags <A> and </A>. *Meta tags* are words located between the <TITLE> and </TITLE> tags (i.e. page title) and in the attributes of the <META> tags of the HTML document. *Full text* refers to all the words in the Web page and lastly, *text passages* are the words located close to the images. An initial experiment is conducted to determine the best size of text passages where 5, 10 and 20 terms before and after an image as well as full text are tested. Text passages of 20 terms before and after an image give the best results. These sources of text are then combined in a Bayesian network model to improve the effectiveness of image retrieval. Text passage when used in combination with description tags give the best retrieval results and poor retrieval is shown when these text sources are used in isolation.

The image contextual information can range from the image caption (i.e. a paragraph of text) to the entire article text body in (Joshi and Liu, 2009). Joshi and Liu (2009) classify HTML tags into either *block* or *style* tags. The block tag element impacts the page layout or the relative positioning of the content whereas the style tag element only affects the visual attributes of the content such as font size or color. The identified block tag is then rendered into a content block on the Web page, which they consider as the article text body (i.e. the main content of the Web page). Then, a linguistics-based semantic similarity algorithm is used to associate a Web image to the article text body. It finds a match for the named entities between the image caption and text from each of the content blocks that make up the article text body. Web images without captions are considered as non-article images.

The remaining works index Web images solely on the surrounding text extracted from the entire Web page content. Gong et al. (2006) subject the image context to stop word removal and stemming processes and then categorize it within three categories – page-oriented text (i.e. text from the title and meta tags), link-oriented text (i.e. text attached to the image tag) and caption-oriented text (i.e. text of the body). Caption-oriented texts which are big in size are further segmented in blocks according to HTML tags considered to be semantic delimiters (<BODY>, <TABLE>, <P>, <UL>, <DIV>, etc.). For each term in a block, a local weight corresponding to its semantic relevance to the Web image is calculated based on its local occurrence using the *tf-idf* weighting model and distance of the block to the image. Then, the overall relevance of a term to a Web image is determined as the sum of all its local weight values multiplied by the corresponding distance factors, thus, attempting to rank relevant terms higher than irrelevant terms.

Leong et al. (2010) improve the work in Leong and Mihalcea (2009) using a linguistics-based approach to automatically annotate Web images from any domain with denotative (i.e. depicted entities or objects) and connotative (i.e. interpreted semantics or ideologies) keywords by relying exclusively on the information drawn from the image contextual information. They consider the entire Web page content as the image contextual information and use knowledge bases (Flickr and Wikipedia) to model and measure the denotative and connotative quality of a word based on two textual features: picturability and salience respectively. The top-*k* words from a ranked word list according to the picturability and salience scores are assigned to an image as keyword annotations.

## 3.2 Fused Systems

In earlier works of loosely-coupled models that utilize both image contextual information and low-level image content (such as colors, textures, shapes…), WebSeer (Frankel et al., 1996), AMORE (Mukherjea et al., 1999), ImageRover (Sclaroff et al., 1999) and iFind (Chen et al., 2001) use textual cues that come from the image filename, ALT attribute within the  tag of the HTML file, link text, title of the HTML page and image surrounding text. The definition of image surrounding text differs between the systems. Sclaroff et al. (1999) define it as 10 words appearing before the  tag (i.e. the Web image) and 20 words appearing after the  tag. Emphasized words in bold and italics, word frequency and word proximity to the image are all taken into account using the Latent Semantic Indexing method. For word proximity, the closer words are ranked higher. Frankel et al. (1996), Mukherjea et al. (1999) and Chen et al. (2001) do not specify a number of words before and after an image, instead they consider a paragraph of text. Frankel et al. (1996) weigh the texts according to the locations in the source Web page of image filename, image caption, ALT attribute, HTML page title and link text, which they just assume to be of importance in the context of image search whereas Chen et al. (2001) simply use a term and document frequency-based weighting scheme (i.e. *tf-idf*) to rank the texts. More recently, Deschacht and Moen (2007) use a paragraph of text (image caption) in a linguistics-based approach to recognize faces and people in news images. A face recognizer tool is used to detect the number of faces/persons present in an image and text discourse analysis techniques, followed by sentence parsing, are applied to detect the salience of named entities extracted from the text. In extending to the detection of other object entities, the visualness of the entity is measured using the WordNet ontology and is then combined with the entity salience in a probabilistic model to discriminate between relevant and irrelevant words.

Smith and Chang's (1997) WebSeek image and video retrieval system utilizes the hyperlink text and Web addresses, i.e. Uniform Resource Locators (URLs), to index multimedia resources on the Web. A well-formed URL follows the convention:

*URL = http://host.site.domain[:port]/[user/][directory/][file[.extension]]*

where […] denotes an optional argument. A hyperlink parser analyzes URLs that contain image or video files to extract the terms from the file and directory names. Terms are manually assessed based on their descriptiveness and clarity; terms that are both descriptive and not ambiguous are considered as key terms. Phrases are also considered and extracted from directory names. Key terms and phrases are then mapped to corresponding subject classes using a manually constructed key term dictionary and organized into a hierarchical taxonomy.

Feng et al. (2004) carry out a study and define the surrounding texts as text sections separated by structural HTML tags such as <TABLE>, <TR>, <TD>, <DIV> and <HR> with a cutoff point at text description length greater than 32 words before and after an image. The texts are then fused together with visual features using a co-training approach. Similarly, Hua et al. (2005) also defined surrounding texts as text sections but relied on the border properties of structural HTML tags as separators instead.

In Cai et al. (2004), the VIPS (Vision-based Page Segmentation) algorithm is implemented to partition a Web page according to its visual presentation into smaller blocks or sections. Only blocks containing at least an image are considered and termed as image blocks. All the textual information within an image block is defined as the image contextual information. Each image is represented using text, visual, and link information. This information is used to cluster or organize the search results. Other image clustering works (Gao et al., 2005; Rege et al., 2008) simply consider the entire page of text excluding the stop words as image contextual information and focus on the use of visual features to improve the system. These works however do not address the fundamental problem of using textual information to annotate images.

As an improvement, data gathered from Web 2.0 websites such as Flickr and Picasa, have been used to enhance the performance of simple concept-based retrieval systems. In particular, these websites put a specific emphasis on tags for indexing and retrieving the multimedia content.

In Olivares et al. (2008), mimicking the bag-of-words indexing technique employed for text retrieval, the authors model an image with a vector of visual words weighted according to the tf-idf weighting scheme. Visual words are obtained after extracting low-level visual features, quantizing and clustering them. An image is then characterized by a histogram of visual words. As far as querying is concerned, a query-by-example framework is implemented. A post-retrieval stage consisting of rank aggregation over the top results with contextual information (image tags, title and description) matching the query is then employed.

However, Yang et al. (2011) pinpoint the insufficiency of tags in providing a complete description of the document content. They therefore propose to enrich tag descriptions with additional visual properties; i.e. color, texture, size, dominance and a given location. For example, the tag "ball" can be enriched with the color description "white", the texture description "striped", the shape characterization "round"… Image regions corresponding to tags are first highlighted through a learning approach and tags are then enriched with corresponding properties of the image regions. These are shown to outperform the empirical performance of social image search, especially when properties are mentioned in a formulated query.

The issue of noisy tags (i.e. tags that inaccurately characterize the multimedia content) has been addressed by Tang et al. [2009] to highlight the semantic content of Web 2.0 images. A sparse graph-based semi-supervised learning framework is used to infer semantic concepts. A training label refinement strategy within the learning framework allows handling the noise in the tags. A concept space in which the relations among concepts are embedded is constructed to assist the inference mechanism.

Larson et al. [2011] propose themselves to enrich the informativeness of image tags, i.e. how related they are to the image content, by combing a measure predicting whether the tags correspond to a physical entity and its associated size (i.e. small, medium or large) as well as statistics of web-mined natural language.

## 4. ANATOMY OF A CONTEXT-BASED WEB IMAGE UNDERSTANDING FRAMEWORK

We present in this section the organization of a context-based Web image understanding framework. Our aim is to characterize the image content with semantic-based descriptors in order to address the image semantic gap. At the basis of our proposal is the notion of semantic facet: abstraction that aims at characterizing the semantic image content over a specific domain of interest.

To begin with, the image and its contextual information need to be extracted from the hosting Web page. The contextual information will then be processed into a set of tokens or index terms. Bearing in mind that the tokens will be characterized semantically, the text processing stage must prepare the tokens for this automatic task.

For automatic classification, the system must know the meaning of the token in order to classify it correctly. The polysemic nature of a word (i.e. a word can have multiple meanings or senses) now becomes an issue which must be addressed, e.g. "bass" (the fish) shall not be classified as a sound property under the *abstract* semantic facet. As such, word sense disambiguation must take place during the text processing stage. We refer to this text processing stage as the natural language processing stage. Only then, the token can be classified, ranked and used as an image index.

Figure 6 illustrates the framework to automatically extract images and their contextual information, perform natural language processing and exploit a knowledge base to classify the image textual context into semantic information. The proposed framework has two main stages:
- The first stage consists of the identification of the relevant image contextual information.
- The second stage deals with the understanding of the Web image contextual information.

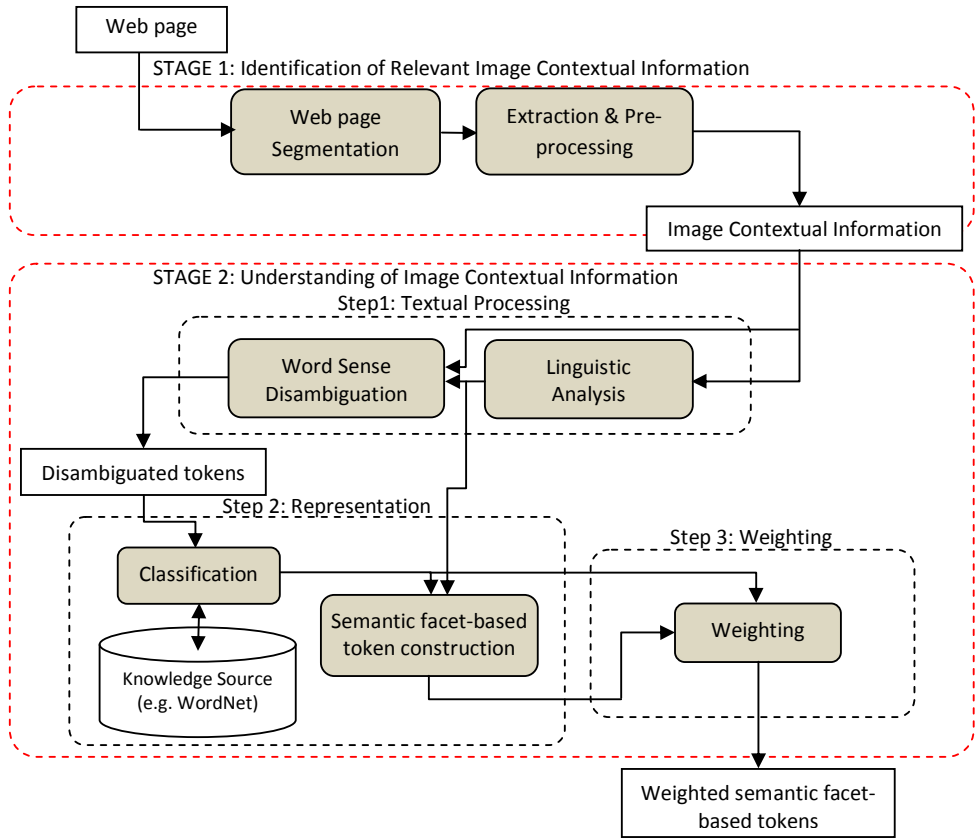

**Figure 6.** Context-based Image Retrieval Framework

## 4.1 Stage 1: Identification of Relevant Image Contextual Information

In the first stage, a segmentation algorithm is required to extract images and their contextual information from a Web page as illustrated in Figure 7. Each Web image and its contextual information are referred to as an image segment (highlighted in the red boxes in Figure 7). After extracting the image segments, a pre-processing module splits the contextual information for an image into visible and hidden texts and performs additional pre-processing steps on the hidden texts. Figure 8 shows the actual extracted text. The boxed texts are the visible texts which are displayed in the Web browser (c.f. Figure 7) and the highlighted texts are the hidden texts. Hidden texts are the values of the HTML tag attributes and they contain much noise.

HTML tag attributes are to provide additional information about the HTML element. Some key functions of these attributes include the specification of style, language and events, unique identifiers and provision of extra information (http://www.w3schools.com). Hence, we can broadly group these attributes into two categories:

Category A: Attributes that contribute to the formatting of the HTML page (e.g. WIDTH, HEIGHT, BORDER, CLASS, SIZE, FACE, CELLPADING, CELLSPACING, etc.)
Category B: Attributes that do not (e.g. SRC, ALT, HREF, TITLE, NAME etc.)

It is often assumed that information coming from page format-related attributes can be discarded. In addition, words that have been stringed together are split and common words, which include file extensions (e.g. jpg, gif, html, cfm, etc.) and HTML related words (e.g. main, index, home, link, etc.), dropped. For example, the character string "http://l.yimg.com/a/i/ww/news/2009/05/24/terminator.jpg" is reduced to "news" and "terminator".

Visible texts, on the other hand, are usually well formed sentences or phrases but face sentence fragmentation issues due to HTML formatting. To address this, there is a need to explicitly join the fragments, based on the Document Object Model (DOM) tree that provides a way for describing and manipulating HTML objects, which can be addressed during or after the segmentation phase. This stage and its instantiation in state-of-the-art works are further discussed in Section 5.1.

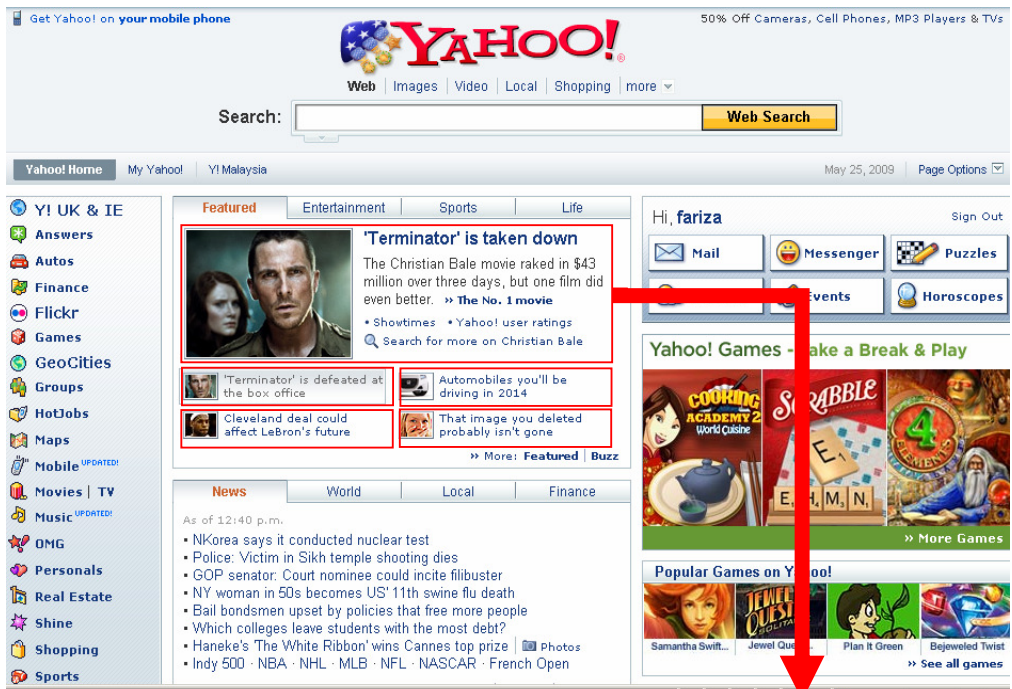

**Figure 7.** The Web page segmentation process

## 4.2 Stage 2: Understanding of Web Image Contextual Information

### 4.2.1 Step 1: Textual Processing.
Here, natural language processes of linguistics analysis and word sense disambiguation are performed on the image contextual information, rather than the common practice of stemming and stop word removal, to identify the words or phrases in their right meaning within a given context. By deciphering the texts, the classification of the image textual context into semantic information and associations between semantic concepts highlighted.

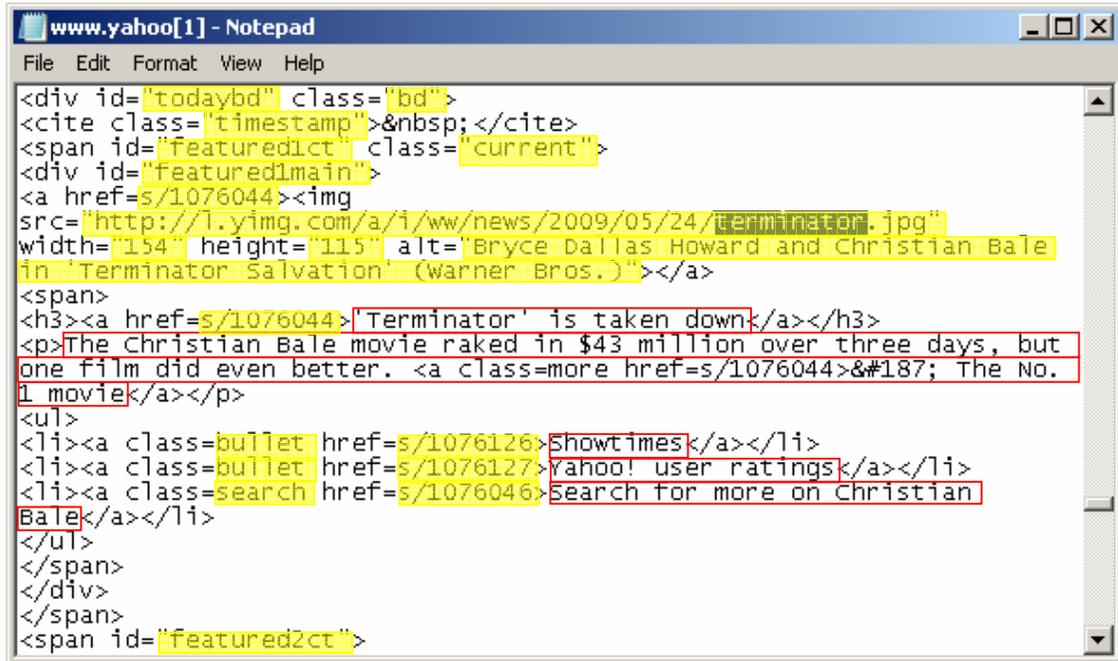

**Figure 8.** Actual image contextual information

State-of-the-art natural language processing tools are employed to tackle the issues of polysemy and non-compositional phrase identification. Linguistic parsers to break up a sentence into components of phrases are utilized, putting them into their functional classes (i.e. semantic roles) and identifying the dependencies between them. For example, a noun phrase can function either as a subject, object, or agent. It also tags each word with its respective part of speech (e.g. nouns, verbs, adjectives, and adverbs) based on the sentence syntax. These classes and dependencies, as well as the part of speech tags, are important information used to identify non-compositional phrases at this stage and perform the semantic characterization in the final stage.

Non-compositional phrases are defined as named entities, collocations, phrasal verbs, idioms, metaphors or acronyms that are treated as single entities with their own meanings. Idioms and metaphors pose some problems when used jointly with images. For example, as far as the idiom "to let the cat out of the bag" is concerned, some Web authors visualize its literal meaning, while others consider it metaphorically.

Next, the tokenization process takes place splitting the contextual information into tokens. Non-compositional phrases are considered as single tokens (e.g. "fountain_pen"). Given a token in its context, its meaning (i.e. sense) shall be identified. The method which gives superior results in practice consists of using a window of surrounding words to disambiguate word senses. For example, *bank* can mean a depository financial institution or a sloping land near a body of water. If it has words *money* or *loan* nearby, it probably is in the financial institution sense; if it has the word *river* nearby, it is probably in the sloping land sense (Pedersen and Kolhatkar, 2009). Similarly, this method can be employed to perform word sense disambiguation and at the end of this stage, each token is attached a sense number and part of speech tag (c.f. Figure 9).

**4.2.2 Step 2: Classification/Semantic Representation.** Once the image contextual information is broken down into tokens with their respective senses and part of speech tags, it is ready for semantic facet-based characterization. This aims at giving some understanding and orderliness to the contextual information where we may find numerous and varying semantic concepts. Hence, for each token, it will be assigned to one or more facets according to its semantics. An external knowledge base is essential to accomplish this. Lexical ontologies, organizing concepts in semantic hierarchies, are used to perform the semantic characterization of tokens. For example, the WordNet ontology is widely diffused in the research community as the knowledge base for the English language to perform automatic concept indexing and classification (Aslandogan et al., 1997; Jin et al., 2005). It consists of the grouping of nouns, verbs, adjectives and adverbs into sets of cognitive synonyms (synsets/lexicalized concepts), each expressing a distinct concept. Pairs of word senses and synsets are connected by lexical and semantic-conceptual relations (or in short, semantic relations) respectively. Lexical relations (e.g. antonymy, pertainymy, etc) connect word senses included in the respective synsets and semantic relations (e.g. hypernymy, hyponymy, meronymy, holonymy, similarity, see also, troponymy, entailment etc.) apply to synsets in their entirety. WordNet structures each word/lexical class (i.e. parts of speech/syntactic categories of nouns, verbs, adjectives and adverbs) separately; one semantic network per class with different semantic relations linking the concepts in a hierarchical organization. The Wordnet 3.0 version contains about 155,000 words organized in over 117,000 synsets. Majority of the words are nouns (76%), followed by adjectives (14%), verbs (7%) and adverbs (3%).

Furthermore, the syntactical dependencies highlighted by linguistic parsers to discover associations between tokens are leveraged for the construction of multifaceted indexes, i.e. semantically coherent combinations of faceted concepts.

### 4.2.3 Step 3: Weighting.
This step involves emphasizing the contribution of some features of the highlighted token to estimate their relevance to the image As such, each token is weighted to estimate its importance to the image. Weighting schemes and their implementation in the state-of-the-art works are detailed in Section 5.

We have presented an overview of the proposed multifaceted semantic concept-based indexing framework. Each stage is described in length in the subsequent sections. In Sections 5 and 6, we discuss the usage of contextual information in the state-of-the-art in accordance to the two stages described in Section 4. We take a close look at issues within each stage and how they are being tackled in related works. Section 5 highlights the processing of the Web image contextual information from its identification to its extraction (Stage 1) and Section 6 covers the second stage on the semantic processing and characterization of the information using natural language processing tools and external knowledge bases to arrive at the faceted image representation.

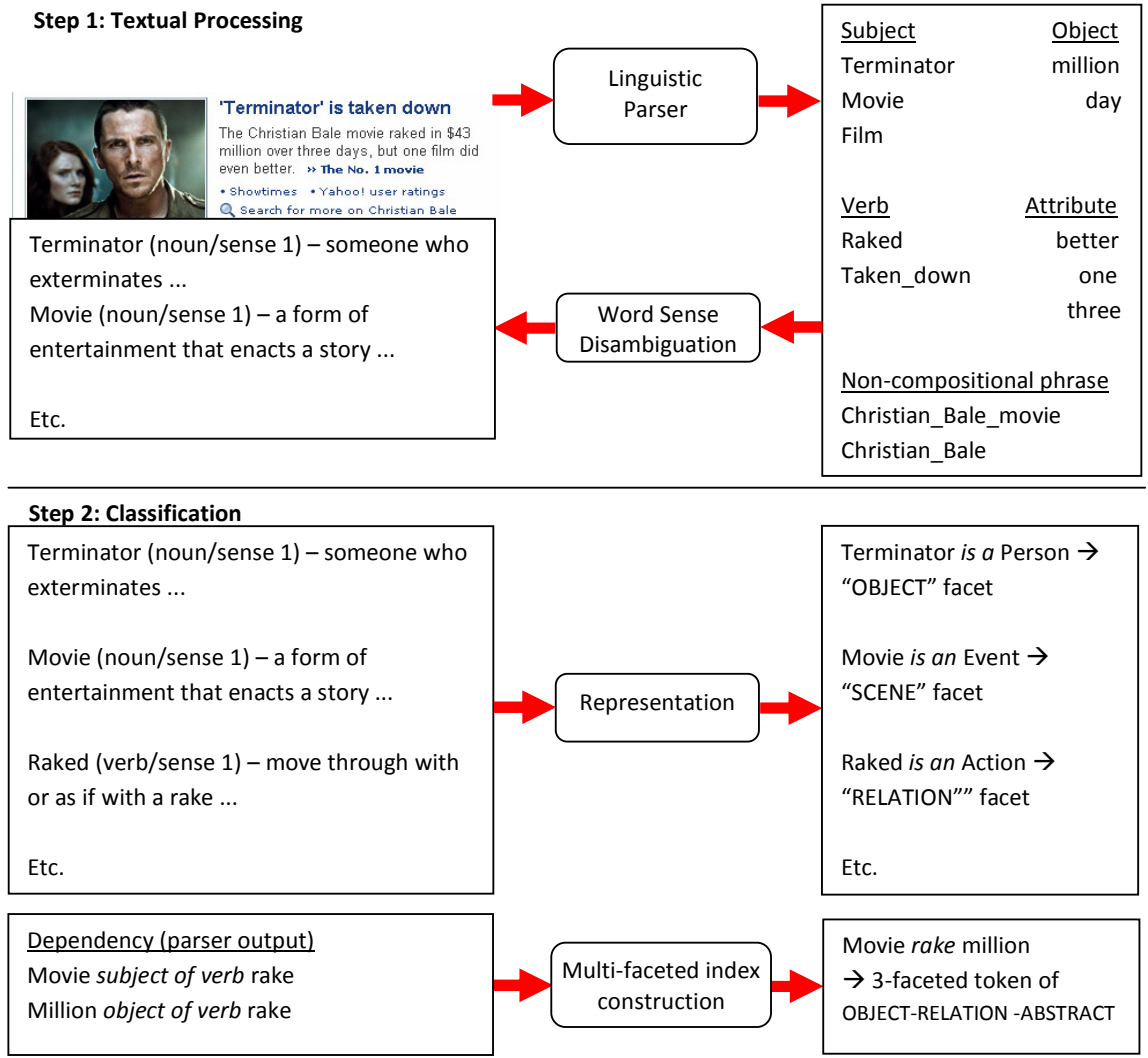

**Figure 9.** Example outputs of Stage 2

# 5. DEFINITION AND AUTOMATIC EXTRACTION OF WEB IMAGE CONTEXTUAL INFORMATION

## 5.1 Highlighting of Relevant Web Image Contextual Information

This section is motivated by the inconsistency of what is considered as the contextual information of a Web image found in the literature. It is clear that no standard work exists as how to associate contextual information relevant to the content of Web images and to the best of our knowledge, only one study on the Web image contextual information that considers the Web users' viewpoints was carried out in the literature (Fauzi & Belkhatir 2010). Varying definitions of image contextual information are found. Existing systems typically extract image contextual information from one or more of the following locations:

- Image filename – text that is located in the SRC attribute of the image tag 
- Image ALT text – text that is located in the ALT attribute of the image tag 
- Page title – text between the <TITLE> tag pair in the header of an HTML document
- Hyperlink (link) text – text between the anchor tags <A> and </A>
- Uniform Resource Locator (URL) - Web addresses that are located either in the HREF attribute value of the anchor tag <A> or in the SRC attribute value of the image tag 
- Surrounding text – text appearing before and after the image, ranging from 10 words, 20 words, a paragraph (caption) or word section to the entire page.

Existing definitions differ due to the diverse designs of Web pages. Earlier Web pages were simpler in terms of structure, style and content. These Web pages in general contain very few images and mostly consist of text. Considering a fixed number or even the entire page of terms as the image contextual information may be suitable for such Web pages. As such, automatic extraction of the image surrounding text is easily achieved via fixed window methods. The definition of relevant image contextual information plays a role in deciding its extraction method (or vice versa as some researchers opt for a window of terms or full text for easy extraction). In Leong et al. (2010) where the entire page of text is considered as image context, a dataset of images from Web pages containing a single image only is constructed (other Web pages are excluded due to their prposed definition of image surrounding text). Current Web pages can be complex with multiple topics and images all packed into a page.

For the "section of text" definition, the challenge lies in the implementation of an automatic extractor that caters to the assorted Web pages. Taking a paragraph of text as the image surrounding text (Deschacht and Moen, 2007; Frankel et al., 1996; Joshi and Liu, 2009; Shen et al., 2000), which is the text that is located within the same HTML tags which contain the  tag, is the simplest method but again limited to Web pages that use this type of Web authoring convention (e.g. Deschacht and Moen (2007) only consider news pages). It is not uncommon to find HTML codes that separate the image and text in different HTML tags and Joshi and Liu (2009) assume such images are non-article images (i.e. advertisements). The research focus has moved towards Web page segmentation techniques to dynamically extract sections containing the images and their surrounding context.

## 5.2 Web information extraction vs. Web page segmentation

There has been considerable research on the general problem of extracting information from Web pages. The latter are considered as semi-structured pages. Most Web information extraction systems look into automating the translation of the semi-structured input Web pages into structured data important to many Web mining and searching tools, e.g. information browsing, meta-search engines, query answering, pattern mining in product comparison, etc.

They rely on the fact that the majority of dynamic Web pages have some forms of underlying templates. Earlier works involve human-intensive hand-crafted extraction rules (Crescenzi and Mecca, 1998; Hammer et al., 1997; Sahuguet and Azavant, 2001). Works in (Arasu and Garcia-Molina, 2003; Crescenzi et al., 2002; Hong et al., 2010; Liu et al., 2010; Wang and Lochovsky, 2003; Zhai and Liu, 2005) focus on the automatic extraction of structured data from Web pages relying on automatically generated templates (Arasu and Garcia-Molina, 2003), wrappers (Crescenzi et al., 2001; Wang and Lochovsky, 2003), tree-matching algorithms (Hong et al., 2010; Zhai and Liu, 2005) and vision-based approaches (Liu et al., 2010) respectively. Further reading on a comparison of Web information extraction systems in terms of task domain, degree of automation and techniques used is found in Chang et al. (2006). We have yet to see Web information extraction techniques being applied to the task of extracting Web images and their textual context. We however discuss three issues potentially impacting the performance of Web information extraction. First, the ambiguity in defining the location of relevant contextual information hampers its automated extraction. Then, the heterogeneity of Web pages in terms of contents and layouts directly affects automated extraction tools which require several similar Web pages to perform satisfactorily. Finally, the required parameter tuning of the automated extraction process is Web page dependent and fastidious.

On the other hand, Web page segmentation is the task of breaking a Web page into regions that appear coherent to a user browsing the Web page. The regions are mainly the actual page content, navigation menus, header and footer, and advertisement banners. For some Web pages, the actual content can be further segmented to smaller sections, for instance in a news page, a user might view the main content as a number of sections with different contents, e.g. World, Local, Business, Entertainment, Sports, etc.

A Web page has been segmented in a number of different approaches for various purposes, mainly for information retrieval applications such as keyword-based Web search, content/information extraction (Spengler and Gallinari, 2009), automatic page adaptation (Chen et al., 2003; Kang et al., 2010), de-duplication detection (Chakrabarti et al., 2008) etc. Most of these Web page segmentation algorithms are typically DOM Tree-based approaches as they allow efficient analysis and manipulation of the Web page content. Pnueli et al. (2009) treat the Web page segmentation problem as a pure image segmentation problem. However, such visual approach is computationally intensive.

Figure 10 shows the example of a Web page that is mapped to a DOM structure from http://www.w3.org. The structure is hierarchical (i.e. tree-like) due to the embedding nature of HTML. Each node of the DOM Tree is an HTML tag element. All text and images that appear on the Web page are leaf nodes of the DOM Tree.

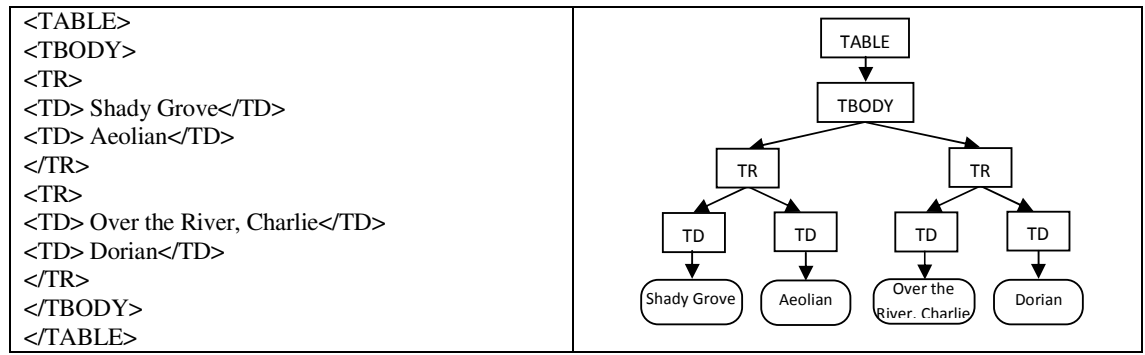

**Figure 10.** An example of a table taken from an HTML document and its graphical representation as a DOM structure

Kao et al. (2005) separate blocks of the DOM sub-trees by comparing the entropies of the terms within the blocks. Chakrabarti et al.'s (2008) meta-heuristic Graph-Theoretic approach casts the DOM tree as a weighted graph, where the weights indicate if two DOM nodes should be placed together or in different segments. Cai et al. (2003) approach the segmentation problem from a computer vision perspective in their VIPS algorithm. Segmentation is performed based on DOM Tree-based heuristics and additional visual cues. An adjustable pre-defined Permitted Degree of Coherence (PDoC) value can be set to achieve different granularities of content structure to cater for different applications. Lastly, Kohlschutter et al. (2008) apply quantitative linguistics strategies.

Cai et al. (2004) and Wang et al. (2004) first demonstrate the application of the VIPS Web page segmentation algorithm (Cai et al. 2003) in image retrieval systems to extract image contextual information. Recent works of He et al. (2007) and Liu et al. (2009) also use the same extraction technique. We anticipate the adjustable PDoC value, which provides the flexibility to control the granularity of resulting blocks from a Web page, to pose problem on Web pages with different block/section sizes on one page.

Alcic and Conrad (2011), Feng et al. (2004), Hua et al. (2005), Joshi and Liu (2009) and Li et al. (2006) segment Web pages specifically to extract Web images and their contextual information. Feng et al. (2004) and Hua et al. (2005) use a similar segmentation approach; a Web page is mainly segmented by identifying structural/separator tags in the DOM Tree. While efficient, the heuristics used above work on limited Web pages, and Feng et al. (2004) fall back on the use of a fixed window size. Li et al. (2006) use visual cues of size and position in their page segmentation algorithm. Visual-based algorithms are known to be computationally expensive and become crucial when processing the large-scale Web. Joshi and Liu (2009) use a linguistic-based approach to extract Web images and the contextual information. They assume that only images with image caption are valid images and the decision to cluster text elements as one block/segment depends only on named entity matching between image captions and other textual elements. A more recent approach is by Alcic and Conrad (2011) who use a clustering-based approach to extract the image context.

## 6 UNDERSTANDING OF WEB IMAGE CONTEXTUAL INFORMATION

The image contextual information upon extraction is processed to be converted into a set of index terms where a term refers to either a word or a phrase. There are two general approaches found in the existing Web image understanding frameworks – standard and linguistics-based processing. We then deal with issues pertaining to the semantic representation of the contextual content and the weighting scheme to assess its importance.

## 6.1 Textual Processing

**6.1.1 Standard Processing.** Most common, as in (Cai et al., 2004; Feng et al., 2004; Gao et al., 2005; Gong et al., 2006; Hua et al., 2005; Quack et al., 2008; Rege et al., 2008; Shen et al., 2000; Wang et al. 2004), the image surrounding text is tokenized and using a stop word list, identified stop words are removed. Stop words are functional words that do not contribute to the content such as prepositions, conjunctions, articles and pronouns. Next, stemming, which is the process of reducing words to their base/root forms, is optionally performed (Feng et al., 2004; Gong et al., 2006; Shen et al., 2000). On one hand, stemming improves retrieval performance by treating words with the same stem as synonyms where documents indexed with the term "car" are returned for a search query of "cars". On the other hand, on top of the general weakness of stemming where unrelated words are inadvertently mapped to the same stem (Anick, 1991), it is arguable that in information retrieval, "car" is not the same as "cars" as the user might have intended to search for documents containing a single car for the former. Finally, the remaining tokens are then considered as the set of image index terms in concept-based systems or as the textual features of the image in multi-modal systems.

In image retrieval systems implementing the Latent Semantic Indexing technique (Google; La Cascia et al., 1998; Sclaroff et al., 1999; Westerveld, 2000), the same text processes are applied with the additional elimination of terms that appear in a single document only for the reason that such terms do not play a role in discovering inter-document relationships.

This standard method of text processing typically breaks text down into single word tokens separated by white spaces and eliminates stop word tokens. Non-compositional phrases are not taken into consideration. In a simple tokenization process (i.e. using white spaces), collocations such as "fountain pen" and "school bus" are simply tokenized into "fountain", "pen", "school" and "bus" respectively although they should be treated as single entities. Consequently, images of a (fountain) pen and a (school) bus are wrongly indexed with the words "fountain" and "school" respectively.

Another issue lies with the current stop word lists where prepositions are included. While for pure text documents, removing prepositions does not change the content or gist of the document; for image documents, prepositions are the words that describe the objects' relative position in an image. The sentence "A bird flying high *above* the house" brings to mind an image of a bird above a house and not of a bird on the garden ground next to a house, pecking for worms. The preposition *above* is crucial and gives the relative position of the bird to the house, similar to the presence of the preposition *on* in the query illustrated in Figure 2.

With all the issues brought about by the simplistic text processes, the focus has shifted to more elaborate linguistics-based processing.

**6.1.2 Linguistics-based Processing.** This approach involves the application of natural language processing, which is the area of research that explores how computers can be used to understand and manipulate natural language text or speech to perform automated tasks such as machine translation, text summarization, elaboration of expert systems and so on. A natural language processing system may start the processing at word level – to determine the morphological structure, nature (such as part-of-speech, meaning) etc. of the word; then proceed at sentence level – to determine the word order, grammar, meaning of the entire sentence, etc.; and finally consider the context and the overall environment or domain. A given word or a sentence may have a specific meaning or connotation in a given context or domain, and may be related to many other words and/or sentences in the given context (Chowdhury, 2003). As the development of natural language processing tools continues to mature, we observe their increasing use in image understanding. Recent works (Deschacht and Moen, 2007; Hua et al., 2005; Leong and Mihalcea, 2009, Leong et al. 2010) implement a simple linguistics-based approach, which is to perform part of speech tagging on the image contextual information (i.e. to detect the syntactic word category such as noun, verb, adjective, adverb, etc.) and then to retain the nouns only as the textual feature. Additionally, Feng and Lapata (2008) consider verbs and adjectives as well. The linguistics-based approach is certainly an improvement over the standard way of text pre-processing. However, retaining nouns or noun phrases alone is insufficient as it is common to use non-noun words to describe (or search for) an image e.g. "sad", "presenting" etc. While Feng and Lapata (2008) did not exclude verbs and adjectives, their work is restricted to the news domain.

The importance of multi-word tokens (i.e. non-compositional phrases) is only realized in these recent years where non-compositional phrases are being addressed. With the linguistics-based approach, tokens could be either single word or multi-word tokens. Both Hua et al. (2005) and Leong et al. (2010) employ an n-gram technique to automatically identify and extract phrases. Hua et al. (2005) rank the possible n-gram phrases based on four phrase features: *visual weight, phrase length*, *weight* and *independence*, from which a regression model is learned to map them into a single salience score. Thus, the textual description of an image is set as the phrases whose salience score are among the top three ranks. Identified phrases are not limited to non-compositional phrases. Leong et al. (2010) rely on an external knowledge source, Wikipedia and consider n-grams found in Wikipedia only. Identification of non-compositional phrases is still an open AI-complete problem (i.e. a task whose solution is at least as hard as the most difficult problems in artificial intelligence) in the area of computational linguistics (Lin, 1999; Katz and Giesbretcht, 2006). Furthermore, there are no comprehensive knowledge sources on non-compositional phrases available yet.

The full processing of textual components would highlight the relationships between words. Simple tokenization will remove all relationships, leaving all tokens unrelated and equal. Performing word-level natural language processing techniques (e.g. part of speech tagging), semantic clustering (Hearst, 2006; Leong and Mihalcea, 2009), or hierarchical taxonomy/topic modeling (Feng and Lapata, 2010) also do not fully address this problem. Semantic clustering and hierarchical taxonomy development mainly tackle the issue of generality or specificity (considering the is-a relationship) between tokens (e.g. *cow* is an *animal*). As such, this particular issue remains unaddressed, it is illustrated in Figure 5 where an image of a cow on a car is returned in the first page of search results for the query "Car hits lorry". The image contextual information consists of: "*CowOnTheCar.jpg*", "*The animal was dropped from a lorry driving in*" and "*funnyanimalsvideo.org*". When the words are in the proper context, we can easily see the relationships between them: the *cow* is on the *car*; the *cow* is the *animal* from the *lorry* and other than the *cow*, the *car* has no association to the *lorry* at all.

### 6.2 Representation of Web Image Contextual Information

The problem of the bag-of-words representation, which is commonly implemented (Cai et al., 2004; Chen et al. 2001; Coelho et al., 2004; Deschacht and Moen, 2007; Frankel el at., 1996; Gong et al., 2006; Leong and Mihalcea, 2009; Leong et al., 2010), has been highlighted in Section 2. An alternative bag-of-phrases representation is proposed by Hua et al. (2005). While this representation addresses the issue of non-compositional phrases and specific queries, it still ignores relations between objects. Shen et al. (2000) address this with their ChainNet model which chains sentences that share at least one same term. The weights are then heuristically obtained and assigned to each lexical chain.

In designing an image understanding model, the representation of image contextual information should be addressed to take into account both specific and complex/compound queries at the same it should reflect the way humans describe images with words.

In the next section, we first justify the need to consider such queries. Then, we explore what a suitable representation of the image contextual information is instead of a bag-of-words (or bag-of-phrases) model. Existing image description models are studied as these models take into account human-related aspects in image understanding and characterization.

### 6.2.1 User Queries for Web Images.
Both general users (Goodrum and Spink, 2001) as well as image professionals (Jörgensen and Jörgensen, 2005) tend to use short queries to search for Web images. However, it is reported in (Goodrum and Spink 2001; Jörgensen and Jörgensen, 2005) that modifications of image queries are frequent. Goodrum and Spink (2001) give examples of query modifications which consist of the addition of terms (called modifiers) such as: older, young, beautiful, animated, with, on etc. These modifiers are typically adjectives and prepositions. Specific attributes of images that are typically named and sought after are: themes, emotions and relationships (Jörgensen and Jörgensen, 2005).

Two recent studies on failed user queries for Web images (Pu, 2008; Chung and Yoon, 2010) show that the illustrated example queries above are not uncommon. The failed queries tend to be unique and specific. Pu (2008) termed the specific queries as *refined* and discussed several refined types broadly classified either as conceptual or perceptual. Conceptually refined queries require an intellectual interpretation or emotive response, such as time or users' preferences while perceptually refined queries indicate direct visual information about an image, such as color. Chung and Yoon (2010) suggest that image understanding may be improved by utilizing associated information and consider in particular the interlacing of multiple image attributes. Earlier on, Armitage and Enser (1997) highlighted the requirements of different levels (facets) of queries, based on the Panofsky/Shatford Model (Shatford, 1986), in their study of user queries addressed to seven picture libraries. The outcomes of these user studies all demonstrate the need to cater for specific/compound/complex/long queries. Next, we describe existing image description models including the Panofsky/Shatford Model.

### 6.2.2 Existing Approaches to Describing Images.
As early as 1962, Erwin Panofsky came up with a theory to structure content descriptions of images in his work titled "Meaning in the Visual Arts". He suggested three levels of semantic categorization of the Renaissance art: the pre-iconographical description (i.e. the non-symbolic or factual subject matter of an image includes generic actions, entities, and entity attributes in an image), the iconographical analysis (i.e. individual or specific entities or actions) and the iconological interpretation (i.e. the symbolic meaning of an image) (Panofsky, 1962).

Shatford (1986) extends Panofsky's theory to all types of images and categorizes the subjects of pictures as "Generic of", "Specific of" or "About" and under each level, four content description facets are highlighted: who, what, when and where. This model is known as the Panofsky/Shatford Model.

Jaimes and Chang (2000) propose a ten-level image descriptor model, based on describing the image with respect to: type/technique, color, texture, shape, generic-object, generic-scene, specific-object, specific-scene, abstract-object and abstract-scene descriptors. The model is organized in a pyramidal structure according to the amount of knowledge required to characterize the visual content. Levels 1-4 are associated to the perceptual or signal information within an image and Levels 5-10 are conceptual and can be viewed as incorporating the Panofsky/Shatford model (i.e. the Generic of, Specific of and About/Abstract descriptors). At each level of the Panofsky/Shatford model, the differentiation between the "Object" and "Scene" concepts is performed. The "Object" concept refers to an individual entity within an image and the "Scene" concept refers to a description about the image as a whole.

The frameworks above are proposed for the purpose of structuring image indexes. There are other existing frameworks that focus on the searching of images (Armitage and Enser, 1997; Eakin, 2002). In these papers, studies are conducted on query collections with the aim to develop a general-purpose categorization of user queries for still and moving visual images. Armitage and Enser (1997) discover the need to incorporate additional attributes to deal with the medium of imagery as there is a problem with the Panofsky/Shatford Model. These attributes are included in Jaimes and Chang's (2000) 10-level pyramidal model. While Armitage and Enser (1997) give an example of image medium, there are other attributes such as the creator/photographer of the image, date of the image, etc. Hollink et al. (2004) put this information into perspective and categorize it as *non-visual*. By focusing on image descriptions, their combined framework is meant for both indexing and searching. Three levels of image descriptors are suggested, i.e. the non-visual, perceptual and concept levels. The concept level is further divided into three sub-levels : general, specific and abstract. Hollink et al.'s (2004) perceptual and concept levels can be compared to the levels highlighted in the Panofsky/Shatford Model and the Jaimes and Chang's (2000) pyramidal model.

We observe that the classification models discussed above generally conform to Panofsky's theory. Hollink et al.'s (2004) proposed framework is more comprehensive compared to the others as their work integrates the different classification methods and considers the users' image characterizations from both the description and search perspectives.

**6.2.3 Classification of Image Descriptors.** Typically, the classification of image descriptors requires human intervention (Armitage and Enser, 1997; Hollink et al., 2004; Meghini et al., 2001). However, Yee et al. (2003) manage to perform a semi-automatic classification of phrasal descriptions attached to the images from the Fine Arts Museum of San Francisco image collection by comparing the words in the descriptions to the higher-level category labels in WordNet. On the other hand, a fully automatic classification is achieved from a content-based perspective only whereby concepts derived from low-level visual features are classified into high-level concepts such as scene concepts, object concepts etc (cf. Datta et al., 2008 for a review of these techniques). The main drawback of content-based techniques is the impossibility of deriving, let alone classifying, abstract concepts. In Jaimes and Chang's (2000) work, abstract concepts are obtained from the image contextual information and manually classified. Automatic extraction of specific information such as a person's name (Aslandogan and Yu, 2000; Deschacht and Moens, 2007; Deschacht and Moens, 2008), geographical location (Kennedy and Naaman, 2008; Toyama et al., 2003; Zheng et al. 2010), event (Simon et al., 2007; Quack et al., 2008; Luo et al., 2008), etc. from the surrounding information of Web images are gaining focus; however, we have not encountered work that automatically classifies the contextual information into multiple concept classes/facets.

**6.3 Weighting**

Chen et al. (2001), Feng et al. (2004), Ortega-Binderberger et al. (2000) and Wang et al. (2004) adopt the *tf-idf* (term frequency - inverse document frequency) weighting scheme to estimate the importance/relevance of the extracted texts with respect to the image. The *tf-idf* weight is a statistical measure used to evaluate how important a specific text is to a document in a collection. The importance increases proportionally to the number of times the text $t$ appears in the document ($tf_t$) but is offset by the frequency of the text in the corpus ($idf_t$).

The $tf_t$ term, given in equation (1), depends on the number of occurrences of the text in a document:

$$tf_t = \frac{n_t}{m} \tag{1}$$

where $n_t$ is the number of times word/token $t$ appears and $m$ is the total number of tokens extracted from the document for normalization to prevent a bias towards larger documents.

The $idf_t$ term is a logarithmic value of the total number of documents divided by the number of documents in the collection that contain the text. Its role is to discriminate between documents for the purpose of scoring and is defined in equation (2):

$$idf_t = \frac{\log N}{df_t} \quad (2)$$

where $N$ is the total number of documents in the corpus and $df_t$ is the number of documents containing the token $t$. The $idf$ of a rare token is high, whereas the $idf$ of a frequent token is low. Then, the *tf-idf* value for token $t$ is given by multiplying equations (1) and (2).

Quack et al. (2008) implement a modified *tf-idf* weighting scheme that adjusts/dampens weights of multiple occurring words per document (local weight) and assigns negative weights to terms that appear in more than half the documents (global weight). Sclaroff et al. (1999) implement the *Log-Entropy* weighting scheme in their latent semantic indexing model.

In ranking the terms as image descriptors, term features other than term frequency are often incorporated in the weighting schemes. Sclaroff et al. (1999) include in their local weight a *term appearance/style weight*, favoring bold, italicized, and bigger terms. Hua et al. (2005) coin the name *Visual Weight*, which is similar to the term style weight. It takes into account colored and hyperlink terms in addition to Sclaroff's et al. (1999) definition.

*Term location/type* is another important feature used (Chen et al., 2001; Frankel et al., 1996; Gong et al., 2005; Sclaroff et al., 1999; Shen et al., 2000). It refers to the different sources of text, e.g. filename is located in the SRC attribute value of the  tag, page title in the header section of an HTML page defined by the <TITLE> tag, image caption is the surrounding text before and/or after an image ranging from 10 words to the entire text in a Web page etc. These locations are ranked and assigned weights based on the assumption that certain locations contain more useful (relevant) information to the image.

A term has also been weighted based on proximity, i.e. *term proximity weighting* which is a lexical distance (Gong et al., 2006; Leong and Mihalcea, 2009; Sclaroff et al., 1999). The position of the words in relation to the image is taken into account, on the assumption that words closer to an image in the HTML document are more related to the image.

Recent attempts explore the use of natural language resources to filter out noise (irrelevant words) (Deschacht and Moens, 2007; Leong and Mihalcea, 2009; Leong et al., 2010). Deschacht and Moen (2007) introduce two textual features to detect person and object entities in news images: *salience* and *visualness*. *Salience* measures the importance of an entity in a text and *visualness* determines the extent to which an entity can be perceived visually. Both measures are combined to compute the probability that an entity is present in the image.

Leong and Mihalcea (2009) also adopt the use of salience and visualness features in their attempt to perform image annotation on datasets from unrestricted domains. On top of the two features, they propose a semantic clustering of words into *semantic clouds* (topics) and select the dominant topic in the text, which is the semantic cloud with the largest set of words as well as the lexical distance (*term proximity*) term feature. In a mixed domain, individually, salience and visualness features do not perform as well as the semantic cloud and lexical distance feature. However, when used in combination, the precision of the system is the best but at the expense of the recall performance. The semantic cloud is promising while the other features require further improvements for images from mixed domains but it should be noted that these images are from Web pages with single images. Such images will have a lot of available textual information to build the semantic clouds. We sum up the various textual features used in the state-of-the-art systems in Table 1.

**Table 1.** A summary of textual features used in various weighting schemes in state-of-the-art image indexing models

| Textual Features | State-of-the-art |
|---|---|
| Term frequency | Feng et al. (2004); Frankel et al. (1996); Gong et al. (2006); Ortega-Binderberger et al. (2000); Wang et al. (2004); Rege et al. (2008); Sclaroff et al. (1999) |
| Term location | Frankel et al. (1996) - filename, image caption, alt, title, hyperlink text, other text. Sclaroff et al. (1999) – alt, title, h1 – h6, b, em, I, strong Shen et al. (2000) – single sentences from image caption, alt, page title, reconstructed from sentences in image caption, image caption. Coelho et al. (2004) Gong et al. (2006) - page-oriented text (TM – page title and meta tag attributes), a link-oriented text (LT – image filename and ALT), and a caption-oriented text (BT – |

| | full text) |
|---|---|
| Term proximity/distance | Gong et al. (2006) – use relative distance of the text block to the image<br>Sclaroff et al. (1999) – use an exponential function that decreases with the distance of the term from the image<br>Leong and Mihalcea (2009) - |
| Visual/appearance/style | Sclaroff et al. (1999); Hua et al. (2005) |
| Salience | Deschacht and Moen (2007); Leong and Mihalcea (2009); Leong et al. (2010) |
| Picturablity/visualness | Deschacht and Moen (2007); Leong and Mihalcea (2009); Leong et al. (2010) |
| Semantic cloud/clusters | Leong et al. (2010) |

From Table 1, *term frequency* is the most basic and frequently used textual feature in weighting schemes. *Term location* is another feature that plays a significant role in ranking terms as well as in image retrieval. However, different assumptions are adopted in different works. This is similar for *term proximity* and *visual style features*. More recent works are looking at *semantic term features*. However, these works are tested on Web image collections of limited size and content diversity. The implemented term feature shall be effective across several semantic domains of Web pages with diverse layouts.

Table 2 summarizes the usage of context-based image understanding techniques in the reviewed state-of-the-art Web-based image retrieval systems according to the identified stages of: i) identification of relevant image contextual information and, ii) semantic understanding of image contextual information.

## 7 CONTRIBUTIONS REVISITED

In this paper, we have specifically reviewed the usage of image contextual information for image understanding, and in particular its application to the Web-based tasks of image indexing and retrieval. Then, we discussed issues associated with the highlighted stages involved when developing context-based Web image understanding frameworks. The stages consist first of the identification and processing of relevant image contextual information and then the semantic understanding of image contextual information. In the latter stage, three steps are identified: i) natural language processing, ii) semantic facet-based conceptual characterization and iii) weighting.

In the first stage of defining and extracting relevant image contextual information, firstly there exist differences of opinions particularly in determining the amount of terms to consider as relevant textual information; 10 to 20 words before and after an image, a paragraph, a section, and the whole document have been considered. Tag attribute values such as image filename, ALT descriptions, URLs etc. are sometimes omitted usually by linguistic-based image annotation models as these text values are grammatically ill-formed and consist of many formatting or script-related codes. Commonly, these definitions are assumed by researchers based on their own observations without taking into account the Web users' viewpoints; and some limit their work to specific domains or to specific layouts (e.g. Deschacht and Moen (2007) and Feng and Lapata (2008) only consider news pages, Leong et al. (2010) build a dataset consisting of Web pages with single images etc.).

A standard definition is desired as it will influence the extraction method. Most importantly, the proposed approaches shall not be restricted to any domain or layout. Hence, the validity of this definition across all domains or styles of Web pages requires verification. A standard definition of the relevant contextual information of Web images shall be formulated by performing tasks such as Web page observation exercises and user studies that put forth a set of hypotheses on the characteristics of the Web image contextual information influencing the design of the proposed image understanding framework. The second are performed for validation purposes, ensuring the proposed hypotheses fulfill user needs.

Current Web page segmentation or Web information extraction tools are also unable to handle the heterogeneous Web as discussed in our review. The simplest method consists of using a fixed window size whereby a predefined number of words before and after the image are extracted as the image contextual information. Although straightforward, this method tends to produce low-level accuracy as texts tend to be associated to the wrong image, for instance, when the image description appears only after the image. And when taking the entire page, the context will contain too much noisy information. Alternatively, Web page segmentation is performed to extract sections containing the images and their relevant context. However, current Web page segmentation techniques perform well for some Web pages and poorly on other Web pages. The following problems are illustrated in our assessment of automated extraction tools: i) the ambiguity in defining the boundary of the contextual information of each image which shall be addressed by Web page observation exercises; ii) the heterogeneity of Web pages – several state-of-the-art segmentation techniques face problems with different Web pages having different content layout and typical Web information extraction tools require at least two similar Web pages for them to be able to extract information; iii) the parameters/modifications required to tune general-purpose algorithms to extract images and their relevant context. Additionally, the performance of segmentation algorithms in terms of time taken to process a Web page, extract images and their corresponding relevant context is important. A fast algorithm would be required to cater to the large and growing number of images of the Web. Thus, we suggest an automatic extraction algorithm that is able to partition the Web page according to the defined boundaries of the contextual information for each image on a Web page. The boundaries shall vary for different Web images and the challenge lies in determining these boundaries and extracting the right amount of information given any Web page without any human intervention.

Standard text processing techniques include tokenization, stemming and stop word removal. These techniques are easy to implement but do not consider various issues when dealing with free text such as non-compositional phrases (i.e. the combination of two or more words which gives a specific meaning), the role of prepositions etc. Individual words that make up a non-compositional phrase have totally different meanings; imagine annotating an image of fountain pen with the token "fountain". Prepositions are an important group of words when it comes to relating entities in images, e.g. "bird on house" is not the same as "bird house". Newer works (Deschacht and Moen, 2007; Leong et al., 2010) have shifted to more sophisticated linguistics-based processing techniques such as part-of-speech (POS) tagging, named entity recognition (NER), word sense disambiguation to overcome the weaknesses of the existing techniques of utilizing contextual information. Their focus is more on named entity recognition as well as the semantic clustering of words. Linguistics-based text processing methods specifically to address the issues highlighted in Section 2. Nevertheless, the performance of linguistic-based techniques is vulnerable to the imperfectness of real world text input (i.e. texts that do not adhere to the grammar/syntax) (Ait-Mokhtar et al., 2002). It will be common to find ungrammatical text in the contextual information, in particular, the text from HTML tag attributes. Hence, both standard and linguistic-based text processing techniques shall be used to produce the set of indexing terms from the image contextual information.

Generally, after being pre-processed, the set of indexing terms is the bag of words used to represent an image. This limits the user queries to simple keyword-based queries. Such systems perform poorly when presented with complex/specific queries. User studies have shown that such queries exist (Armitage and Enser, 1997; Chung and Yoon, 2010; Pu, 2008) and inclusion of specific and multiple image descriptors is inherent to the way people describe images (Jaimes and Chang, 2000; Hollink et al., 2004; Panofsky, 1962; Shatford, 1986). Furthermore, Yee et al. (2003) have shown in a user study-based evaluation of their proposed faceted metadata search system that 90% of the study participants strongly prefer the category-based approach.

We reason that Web image contextual information can be regarded as a set of image descriptors from the perspective of Web page authoring where the basic principle is that an author of a Web page is likely to put an image on a page together with his/her description that, from his/her point of view, is relevant to the image. Based on this assumption, we are bound to find the classes of image descriptors as proposed in (Jaimes and Chang, 2000; Hollink et al., 2004; Panofsky, 1962; Shatford, 1986). Therefore, we put forward the idea of organizing the set of indexing terms from the image contextual information into multiple classes (which we also refer to as facets and used interchangeably throughout the text), giving users the flexibility to choose which classes (facets) of an image they wish to query (e.g. querying for an abstract concept of the image) as well as giving researchers or system developers of hybrid image retrieval systems the flexibility to select which classes of the contextual image they wish to utilize as textual feature. This process of organizing into the various classes must be automatic since we are dealing with the Web. There is also the relationship between indexing terms that must be modeled to address the issues highlighted in Figures 4 and 5.

Existing image descriptor models presented in Section 5.3 lay the foundation of a multifaceted semantic indexing model. We are not proposing new concept classes; instead, we are making use of predefined high-level concept classes from the existing image descriptor models that are distinctive with the possibility of being classified automatically. Panofsky (1962) proposes three levels – *pre-iconographical* (generic non-symbolic subject matter), *iconographical* (specific non-symbolic subject matter) and *iconological* (symbolic subject matter). Jaimes and Chang (2000) propose *perceptual* and *conceptual* levels. Hollink et al. (2004) propose *non-visual*, *perceptual* and *conceptual* levels. Under the *conceptual* level, Jaimes and Chang (2000) propose the *generic-object*, *generic-scene*, *specific-object*, *specific-scene*, *abstract-object* and *abstract-scene* descriptors. Hollink et al. (2004) further partition the *conceptual* level into three sub-levels of *generic*, *specific* and *abstract*. These sub-levels can be further specified as *scene* and *object*, and subsequently *event*, *time*, *place* and *relation*. Therefore, the faceted model to characterize the contextual information semantically shall highlight a *perceptual* facet to describe the signal information (i.e. colors, textures, shapes…), an *abstract* facet for all symbolic descriptors of the image, a *scene* facet for global image descriptors, an *object* facet for specific entities and a *relational* facet to model the relations between entities. *Event* and *place* are typically *scene* concepts and *time* is an *abstract* concept. As such, an image descriptor model of five classes (facets) (i.e. *perceptual/signal*, *abstract*, *scene*, *object* and *relation/relational*) shall be considered. The continuous advancement in linguistics-based tools and available external knowledge bases shall enable researchers in the domain to fully automate the concept classification process. The *generic* and s*pecific* descriptors, which are broad in essence, should be used in conjunction with a first step of classification into one the proposed facets.

Lastly, to reduce noise, the resultant set of terms characterizing the image content (obtained after the image contextual information is pre-processed) are weighted to rank relevant terms higher than irrelevant terms. Several textual features have been proposed: term frequency, location, proximity, appearance, salience, visualness, and semantic cloud (c.f. Table 1). Term frequency is generally accepted and the basic feature used in all weighting schemes. Term location plays a significant role in ranking term relevance in image retrieval systems but locations are prioritized differently across different works, similarly for term proximity and appearance. The newly proposed semantic term features, i.e. salience, visualness and semantic cloud are promising indeed but are yet to be tested on images from diverse Web page sets. These features shall be observed and quantized in a typical Web page observation exercise.

**Table 2.** A summary of the usage of image contextual information in the reviewed state-of-the-art systems

| | State-of-the-art | Image Contextual Information | | | | | | | Extraction | Text Pre-processing | Textual features used in weighting | Representation |
|---|---|---|---|---|---|---|---|---|---|---|---|---|
| | | File name | ALT | Page title | Meta tag | Link text | URL | Surrounding text | | | | |
| Text-based | Shen et al. (2000) | X | X | X | | | | Paragraph | Specific HTML Tag | Standard | Location | ChainNet |
| | Coelho et al. (2004) | X | X | X | X | X | | 40 terms (20 and 20) | Fixed window | Not known | Location | Bag of words |
| | Gong et al. (2006) | X | X | X | X | | | Full text | All text | Standard | Frequency; proximity | Bag of words |
| | Joshi&Liu (2009) | | | | | | | Paragraph | Linguistic-based Segmentation | n/a | n/a | n/a |
| | Leong and Mihalcea (2009) | | | | | | | Full text | All text | Linguistics | Semantic cloud; proximity; picturabily; salience | Bag of words |
| | Leong et al. (2010) | | | | | | | Full text | All text | Linguistics | Salience; picturability | Bag of words |
| Fused Systems | WebSeer (Frankel et al 1996) | X | X | X | | X | | Paragraph | Specific HTML Tag | Standard | Frequency; location | Bag of words |
| | ImageRover (Sclaroff et al 1999) | X | X | X | | X | | 30 terms (10 and 20) | Fixed window | Standard | Frequency; location; proximity | Bag of words |
| | iFInd (Chen et al 2001) | X | X | X | | X | | Paragraph | Specific HTML Tag | Standard | Frequency | Bag of words |
| | AMORE (Mukherjea et al 1999) | X | X | X | | X | X | Paragraph | Specific HTML Tag | Standard | Frequency; location | Bag of words |
| | WebSeek (Smith and Chang 1997) | | | | | | X | | n/a | URL-specific | n/a | Topic-term |
| | Feng et al. (2004) | X | X | X | | | | Section | DOM Tree-based Segmentation | Standard | Frequency | Bag of words |
| | Cai et al. (2004) | X | X | X | | | X | Section | Visual-based Segmentation | Standard | n/a | Bag of words |
| | Hua et al. (2005) | X | | | | | | Section | DOM Tree-based Segmentation | Linguistics | Frequency; Visual | Bag of phrases |
| | Gao et el. (2005) | | | | | | | Full text | All text | Standard | Frequency | Bag of words |
| | Deschacht and Moen (2007) | | | | | | | Paragraph | Specific HTML Tag | Linguistics | Salience; visualness | Bag of words |
| | Rege et al. (2008) | | | | | | | Full text | All text | Standard | Frequency | Bag of words |